\documentclass[10pt,oneside]{amsart}
\usepackage{adjustbox}

\usepackage{tikz}
\usetikzlibrary{automata, positioning, arrows}

      \usepackage{amssymb}
\usepackage{amsfonts,amstext}
\usepackage{graphicx}
\usepackage{url}
\usepackage{amsmath,amstext,amssymb,amscd}
\usepackage{verbatim} 
\usepackage{mathtools}
      \usepackage{tabularx,booktabs}
      \usepackage{multicol}

\usepackage{amssymb,latexsym}
\usepackage{amsmath}
\usepackage{amsthm}
\usepackage{amssymb}
\usepackage{makecell}

\theoremstyle{remark}

\theoremstyle{definition}

\usepackage{amsthm}
\allowdisplaybreaks

\usepackage{multicol}

      \makeatletter
      \def\@setcopyright{}
      \def\serieslogo@{}
     
      \makeatother

\begin{document}

\author{Adam Graham-Squire}
\address{Adam Graham-Squire, Department of Mathematical Sciences, High Point University, 1 University Parkway, High Point, NC, 27268}
\email{agrahams@highpoint.edu}

\author{David McCune}
\address{David McCune, Department of Mathematics and Data Science, William Jewell College, 500 College Hill, Liberty, MO, 64068-1896}
\email{mccuned@william.jewell.edu}

\title[A Mathematical Analysis of the 2022 Alaska Special Election for US House]{A Mathematical Analysis of the 2022 Alaska Special Election for US House}



\maketitle

In the aftermath of the August 2022 Special Election for US House in Alaska, which used ranked choice voting (RCV) to elect the winner, losing candidate Sarah Palin called RCV ``crazy,'' ``convoluted,'' and ``confusing.'' These sentiments were echoed by some of her political allies. Senator Tom Cotton from Arkansas was the most blunt: ``Ranked choice voting is a scam to rig elections,'' he stated. These comments raise the question: is RCV some kind of crazy scam to rig elections? The short answer is: No. (The long answer is: Nooooooo.) However, RCV does have its drawbacks, which were on full display in this election. In this article we'll discuss the deficiencies of RCV from the mathematical perspective of voting theory, using the Alaska House election as a case study. We note that most of the findings of this article were independently and concurrently discovered by Navr\`{a}til and Smith (see \url{https://litarvan.substack.com/p/when-mess-explodes-the-irv-election}).

\section*{What is RCV?}

In an RCV election, voters submit ballots ranking the candidates in order of preference. The August 2022 Special Election for US House in Alaska (which we'll shorten to ``AK election'') contained the three (not write-in) candidates Nick Begich, Sarah Palin, and Mary Peltola, and voters cast preference ballots with a ranking of these candidates. For example, a voter could cast a ballot ranking Begich as their first choice, Palin as their second choice, and Peltola as their third choice. We'll use the notation $\succ$ to denote when a voter preferences one candidate to another, and thus such a ballot is denoted Begich $\succ$ Palin $\succ$ Peltola. Voters were not required to submit a complete preference ranking, and so could cast a ballot simply ranking Begich as their first choice and leave the other rankings blank. 

Once all of the ballots are cast they are aggregated into a \emph{preference profile}, which shows the number of each type of cast ballot. Table \ref{preference_profile} shows the preference profile of the AK election, where for convenience we combine ballots of the form $A\succ B \succ C$ with ballots of the form $A\succ B$, which has no effect on the RCV winner. The profile also disregards write-in candidates. The number 27053 denotes that 27053 voters cast a ballot of the form Begich $\succ$ Palin $\succ$ Peltola; the other numbers convey similar information about the number of voters who cast the corresponding type of ballot.

\begin{table}[]
  \centering

  \begin{adjustbox}{width=\textwidth}

\begin{tabular}{l|ccccccccc}
Number of Voters & 27053   &15467&11290 & 34049& 3652 &21272&47407&4645&23747\\
\hline
1st choice & Begich & Begich & Begich & Palin & Palin &Palin&Peltola &Peltola &Peltola\\
2nd choice & Palin & Peltola & $-$ & Begich &Peltola &$-$&Begich&Palin&$-$\\
3rd choice & Peltola & Palin & $-$ & Peltola & Begich &$-$&Palin&Begich&$-$\\
\end{tabular}

 \end{adjustbox}
  \caption{The August 2022 Alaska Special Election for US House.}
  \label{preference_profile}
\end{table}

The raw ballot data used to make this preference profile is available from \url{https://www.elections.alaska.gov/election-results/}. We note that there is some ambiguity about how the elections office processed the ballots; we processed the ballots in such a way that our numbers match the vote total information that is publicly available on the Alaska Elections Division webpage. It's possible that the numbers in the last six columns do not exactly match what would be published by the Alaska elections office, but based on our analysis of how ballots were processed we are confident in this profile, and any mismatch in vote totals would be tiny and not affect the conclusions of the article.

To choose a winner of an election given a preference profile, RCV proceeds in a series of rounds. In each round, a candidate's first-place votes are counted; a candidate with a majority of the votes is declared the winner. If no candidate achieves  a majority then the candidate with the fewest first-place votes is eliminated and their votes are transferred to the next candidate on their ballots who has not previously been eliminated. The method continues in this fashion until a candidate achieves a majority of the remaining votes.

In the AK election, the first-place votes for Begich, Palin, and Peltola are 53810, 58973, and 75799, respectively, all of which fall short of a majority. Begich receives the fewest first-place votes and is therefore eliminated, and 27053 of his votes are transferred to Palin while 15467 of his votes are transferred to Peltola. The remaining 11290 votes are said to be \emph{exhausted} and removed from the election. After the vote transfer Peltola achieves a majority of the remaining votes with 91266 votes to Palin's 86026, and Peltola is declared the winner.

Now that we understand the mechanics of RCV we turn to some of the strange features of this AK election, beginning with an analysis of the Condorcet dynamics.

\section*{Condorcet Winners and Losers}

A \emph{Condorcet winner} (respectively \emph{Condorcet loser}) of an election is a candidate who wins (respectively loses) all of their pairwise, head-to-head matchups against other candidates. In the AK election, note that $27053+15467+11290+34049=87859$ voters prefer Begich to Peltola while $3652+47407+4645+23747=79451$ voters prefer Peltola to Begich, and thus Begich wins a head-to-head matchup against Peltola. Similarly, 101217 voters prefer Begich to Palin while only 63618 voters prefer Palin to Begich. Since Begich wins both of his head-to-head matchups against the other two candidates, he is the Condorcet winner of the election. Much of voting theory has focused on the study of Condorcet winners because they are considered ``good'' or ``strong'' candidates, candidates who deserve to win the given election. One could argue that Peltola doesn't make sense as an election winner because she loses head-to-head to Begich. 

The AK election provides an example of an RCV election where the RCV and Condorcet winners disagree. This is a rare occurrence for American RCV elections: as far as we know, there has been only one other documented such case, which occurred in the 2009 mayoral election in Burlington, VT \cite{MM2}.  The Condorcet dynamics of this election are an outlier in the landscape of American RCV elections.

Note that the AK election also contains a Condorcet loser: Sarah Palin. Compared head-to-head against the other candidates, voters prefer anyone else. In fact, Palin is not just a Condorcet loser, she is also a spoiler candidate, which leads us to:

\section*{The Spoiler Effect}

Proponents of RCV often argue for its use by claiming that RCV will eliminate, or almost eliminate, the ``Spoiler Effect.'' This term has different definitions; for this paper we'll choose one of the more common definitions, which states that the \emph{spoiler effect} has occurred in an election if the removal of one of the losing candidates causes the winner of the election to change. In this case, the losing candidate is called a \emph{spoiler candidate}. The AK election demonstrates that the spoiler effect can occur under RCV: note that the removal of Palin would change the winner from Peltola to Begich, since Begich wins head-to-head against Peltola. Thus Palin is a spoiler candidate, ``spoiling'' a victory for Begich.

On the one hand, this doesn't seem like such a strange occurrence. It seems natural that the removal of a candidate can shake up the dynamics of the race enough to cause a change in the winner. On the other hand, consider the following story about the behavior of this electorate. If we offer this electorate the options of Begich or Peltola, the electorate chooses Begich by a pretty wide margin. Suppose we then say to the electorate: ``We forgot to mention, there's a actually a third option Palin, would you like to choose her?'' The electorate essentially responds, ``We most definitely will not choose Palin (who is a Condorcet loser), but since you mentioned her, we'll choose Peltola instead.'' If the electorate likes Begich more than Peltola, how can the introduction of a candidate that they like even less than Peltola cause them to change their minds and choose Peltola instead? Of course, this story uses intuition to make a case that the spoiler effect is bad, but the story is compelling.

The spoiler effect under RCV is rarely observed in American RCV elections. There have only been two previously documented cases: the previously mentioned election in Burlington, and a 2021 city council race in Minneapolis, MN \cite{MW}.

\section*{Monotonicity and No Show Paradoxes}

If you were a candidate in an election, would you prefer more support or less support from voters? The question seems non-sensical: obviously, you would prefer more support. In the AK election, though, in some sense Peltola won because she did not receive more support from voters. To see this, suppose that 6000 of the voters who voted just for Palin were persuaded that Peltola is the best candidate and cast the ballot Peltola $\succ$ Palin instead. What effect should this have on the RCV winner of the election? The sensible answer is that there should be no effect: Peltola won the original election, and giving her more support should only cement that win. However, this extra support would actually cost her the election: with these 6000 voters now listing Peltola first, Palin receives the fewest first-place votes and is eliminated from the election first, causing Peltola to face Begich in the last round of the election. The reader can check that even with the additional support, Peltola still does not have enough votes to defeat Begich head-to-head. This is an example of a \emph{monotonicity paradox}, where there exists a subset of voters such that if the winner were to gain support from these voters then the winner would lose. If Peltola had done a better job reaching out to Palin voters, it would have cost her the election.

Monotonicity paradoxes receive attention in the voting theory literature because they are interesting pathologies of RCV-type election procedures, and should be avoided if possible. Out of hundreds of American RCV elections this AK election is only the third to demonstrate such a paradox \cite{FT},\cite{GZ},\cite{MM1},\cite{ON}, and thus such paradoxes seem to occur rarely in practice. (We note that the actual number of such paradoxes may be higher, especially given that ballot data for some RCV elections is not publicly available.)

This AK election demonstrates another paradox which is closely related to a monotonicity paradox. Suppose that 6000 voters who cast the ballot Palin $\succ$ Begich $\succ$ Peltola had not cast their ballots, choosing instead to abstain from the election. What affect should this abstention have on the winner? The sensible answer is that Peltola should still win: she is ranked last on these ballots and thus their removal does not cost her any support; to the contrary, it seems like the removal of these ballots should help cement Peltola's victory because these 6000 voters are supporters of Palin and Begich. However, if these ballots are removed from the election then Peltola would lose, because Palin would be eliminated in the first round and, as above, Peltola would not have enough support to overcome Begich. This is an example of a \emph{no-show paradox}, where there exists a subset of voters who could create a better electoral outcome for themselves by not casting their ballots. These 6000 voters prefer Begich to Peltola, but their participation in the election causes Begich to lose to Peltola.

No-show paradoxes seem to occur even more rarely in actual elections than monotonicity paradoxes. This AK election is the first documented election to demonstrate a no-show paradox in American RCV elections, at least based on ballot data that is publicly available \cite{GZ}.

\section*{So What's the Deal with the AK Election?}

Why does this particular election seem to display so many of the mathematical weaknesses of RCV? The reason seems to be that the election is very ``close.''  That is, the electorate as a whole is conflicted about whether Begich or Peltola should be the winner. As previous voting theory literature suggests, outcomes such as monotonicity paradoxes are much more likely to occur in closely contested elections \cite{GM}, \cite{M}.

Additionally, this election has unusual political dynamics. Begich and Palin are both Republicans while Peltola is a Democrat, leading Begich and Palin to split the Republican vote in a state that traditionally elects a Republican to the US House. Furthermore, Sarah Palin is a unique House candidate: she has a sizable national profile due to her previous run as a Republican vice presidential candidate, and her past behavior has made her a particularly polarizing figure. Thus, even though a majority of Alaskans seemingly would prefer to elect a Republican, a majority would prefer not to elect Palin. Finally, this election was supposed to be a four-candidate contest including independent candidate Al Gross,  but Gross withdrew before the election. In a close election with such peculiar political dynamics, it is not surprising that we see oddities such as the RCV winner differing from the Condorcet winner.

Does this AK election demonstrate that RCV is a bad voting method? If you are especially offended by monotonicity paradoxes then the answer might be Yes, and this is a reasonable position. However, we have known since the Nobel Prize-winning work of Kenneth Arrow that all voting methods which use preference ballots have weaknesses, so there is no perfect election procedure we can use in place of RCV. In a hotly contested race with unusual political dynamics it is possible that two or more candidates have a strong case to be declared the winner, regardless of what voting method we use. In this sense, any voting method could be called a ``scam'' by supporters of a losing candidate. For example, if Begich were crowned the winner because of his status as Condorcet winner, supporters of Peltola would undoubtedly protest such an outcome because she received the most first-place votes and also wins under RCV. This AK election happens to demonstrate many of the weaknesses of RCV but these weaknesses are rarely documented in American RCV elections.  A reasonable opinion, therefore, is that this AK election does not necessarily invalidate the use of RCV.

\end{document}